\documentclass[twocolumn]{aastex62}     

\usepackage{amsmath}
\usepackage{booktabs}
\received{January 1, 2018}
\revised{January 7, 2018}
\accepted{\today}
\submitjournal{ApJ}

\shorttitle{Bayesian rotation inversion of KIC 11145123}
\shortauthors{Hatta et al.}

\turnoffedit

\begin{document}

\title{Bayesian rotation inversion of KIC 11145123}

\correspondingauthor{Yoshiki Hatta}
\email{yoshiki.hatta@grad.nao.ac.jp}


\author[0000-0003-0747-8835]{Yoshiki Hatta}
\affiliation{Department of Astronomical Science, School of Physical Sciences, SOKENDAI\\
2-21-1 Osawa, Mitaka, Tokyo 181-8588, Japan}
\affiliation{National Astronomical Observatory of Japan \\
2-21-1 Osawa, Mitaka, Tokyo 181-8588, Japan}

\author[0000-0001-6583-2594]{Takashi Sekii}
\affiliation{Department of Astronomical Science, School of Physical Sciences, SOKENDAI\\
2-21-1 Osawa, Mitaka, Tokyo 181-8588, Japan}
\affiliation{National Astronomical Observatory of Japan \\
2-21-1 Osawa, Mitaka, Tokyo 181-8588, Japan}


\author[0000-0001-9405-5552]{Othman Benomar}
\affiliation{Department of Astronomical Science, School of Physical Sciences, SOKENDAI\\
2-21-1 Osawa, Mitaka, Tokyo 181-8588, Japan}
\affiliation{National Astronomical Observatory of Japan \\
2-21-1 Osawa, Mitaka, Tokyo 181-8588, Japan}
\affiliation{Center for Space Science, New York University Abu Dhabi \\
P.O. Box 129188, Abu Dhabi, UAE}

\author[0000-0001-9430-001X]{Masao Takata}
\affiliation{Department of Astronomy, School of Science, The University of Tokyo\\
Bunkyou-ku, Tokyo 113-0033, Japan}




\begin{abstract}
A scheme of Bayesian rotation inversion, which allows us to 
compute the probability of a model of a stellar rotational profile, is developed. 
The validation of the scheme with simple rotational profiles and 
the corresponding sets of artificially generated rotational shifts has been successfully carried out, 
and we can correctly distinguish the (right) rotational model, prepared beforehand for generating the 
artificial rotational shifts, with the other (wrong) rotational model. 
The Bayesian scheme is applied to \edit1{a $\gamma$ Dor--$\delta$ Sct type hybrid star}, KIC 11145123, 
leading to a result that the convective core of the star might be rotating much faster ($\sim10$ times faster) 
than the other regions of the star. 
The result is \edit1{consistent with }
that previously suggested by \citet{Hatta2019} based on a 3--zone modeling, 
further strengthening their argument from a Bayesian point of view. 

\end{abstract}



\section{Introduction} \label{sec:intro}
Stellar internal rotation plays an essential role in 
a variety of stellar internal dynamics \citep[e.g.][]{Maeder_text} such as 
generation of magnetic fields via the dynamo mechanism and 
transportation of chemical elements caused by rotationally induced mixing. 
It is therefore of great importance for us to study stellar internal rotation theoretically and observationally, 
the latter of which, in particular, has recently become feasible thanks to 
the establishment of asteroseismology \citep[e.g.][]{Aerts_text}
brought about by high-precision photometric observations by spacecrafts such as Kepler \citep{Kepler} and TESS \citep{TESS}, 
leading to numerous asteroseismic inferences on internal rotation of various types of stars 
including solar-like stars \edit1{\citep[e.g.][]{Benomar2015,Benomar2018,Schunker2016a,Schunker2016b}}, early-type main-sequence stars 
\edit1{\citep[e.g.][]{Kurtz2014,Saio2015,Schmid_Aerts2016,Papics2017,Christophe2018,Ouazzani2018,Li2020}}, and evolved stars 
\edit1{\citep[e.g.][]{Beck2012,Mosser2012,DiMauro2018,Deheuvels2020}}. 


The current understanding of internal rotation of main-sequence stars 
is summarized by \citet{Aerts2019} 
that almost all the main-sequence stars investigated so far are 
exhibiting nearly rigid rotation throughout them, which has not been expected 
based on previous hydrodynamical numerical simulations of angular momentum transfer inside stars \edit1{\citep[e.g.][]{Tayar2013,Eggenberger2017}}. 
To fill the gap between the observation and the theory, 
several mechanisms of angular momentum transfer by, for instance, internal gravity waves or magnetic fields, 
have been proposed 
\edit1{\citep[e.g.][]{Cantiello2014,Rogers2015,Fuller2019}}. 
Thus, asteroseismic studies have definitely been contributing in propelling understanding of the stellar rotation. 


Interestingly, there are also a few asteroseismic researches suggesting 
the existence of rotational velocity \edit1{gradient} 
inside stars \citep[e.g.][]{Benomar2018}; 
stars are rotating rigidly throughout most of the interiors, but not completely. 
One of such stars for which rotational velocity shear inside 
has been suggested is KIC 11145123 \citep[][]{Hatta2019}, 
which is a $\gamma$ Dor--$\delta$ Sct type hybrid star \citep{Bradley2015} and 
has been actively studied based on its well-resolved frequency splitting for pressure (p), gravity (g), and mixed modes, 
revealing the evolutionary stage \edit1{\citep{Kurtz2014,Hatta2021}}, the asphericity \citep{Gizon2016},
 and the internal rotation \citep{Kurtz2014,Hatta2019}. 

\citet{Hatta2019}'s primary focus was on investigating the latitudinally differential rotation of the star, but, 
by carefully checking behaviors of estimates obtained via rotation inversion, 
they have found a hint that the convective core of the star might be rotating $5$--$6$ times 
faster than the other regions of the star. 
Although the suggestion of the fast-core rotation might appear to be incompatible with the current understanding that 
main-sequence stars are rotating almost rigidly, it is actually not the case. 
Previous asteroseismic studies, especially those focusing on early-type main-sequence stars with the convective core and the 
broad radiative region above, have utilized high-order g modes, 
which \edit1{are not established in} the convective core, to infer rotation rates in the deep region; 
what they have estimated are rotation rates in the deep radiative regions. 
In contrast, \citet{Hatta2019} have used mixed modes, which have finite sensitivity, though small, 
inside the convective core, enabling them to obtain a hint of the rotation rate 
of the convective core located beneath the radiative region. 
They have actually confirmed that the star is rotating almost rigidly throughout the radiative region 
and that the convective core is the exception. 

\citet{Saio2021} provide another asteroseismic study which have inferred convective-core rotation of 
fast-rotating $\gamma$ Dor stars by fitting characteristic dips in observed g-mode period spacings, 
caused by the coupling between inertia modes and high-order g modes. 
Though they have found no hint of rotational velocity shears (between 
the convective core and the radiative region above) 
among their targets, 
comparison between the two studies should be helpful for putting further constraints 
on theoretical calculations of angular momentum transfer inside stars.  


In this paper, we would like to take a further step in terms of \edit1{investigating} 
convective-core rotation of stars based on a newly developed scheme of Bayesian rotation inversion 
which enables us to compute probabilities of models of rotational profiles 
based on the so-called global likelihood \citep[e.g.][]{Gregory2005}. 
The first goal is to present the scheme of Bayesian rotation inversion, and the 
second goal is to apply the scheme for KIC 11145123 and to examine if the 
fast-convective-core rotation is obtained or not. 

The structure of the paper is as follows. 
In Section \ref{sec:2}, the mathematical formulations of the Bayesian rotation inversion is presented 
after a brief introduction to asteroseismic rotation inversion, which is based on the perturbative approach. 
Note that the perturbative approach is justified for inferring the internal rotation of KIC 11145123 because of the fact that 
the star is a slow-rotator with the rotation period of about $100 \, d$, which is much longer than 
the dynamical timescale as well as the oscillation periods of the star of a few hours. 
In Section \ref{sec:3}, validation of the developed scheme is carried out 
with simple artificial datasets. 
We apply the Bayesian scheme to KIC 11145123 in Section \ref{sec:4} where 
the basic properties of the star, the comparison of modeled rotational profiles, and results finally obtained have been featured. 
We lastly 
give a summary in Section \ref{sec:5}. 

Finally, we have a note on reference stellar models used in this paper. 
As reference stellar models for rotation inversion, we firstly chose two models; 
one is a model constructed by \citet{Kurtz2014} assuming single-star evolution, 
which was used in \citet{Hatta2019}, and the other 
is a non-standard model of the star constructed by \edit1{\citet{Hatta2021}} taking effects of some interactions with other 
stars into account. 
\edit1{We}, however, \edit1{did not see any} significant differences 
in results of rotation inversion no matter which model \edit1{was} taken as a reference model, 
and thus, we \edit1{will present the results obtained based on} 
the newer model computed by \edit1{\citet{Hatta2021}} in this study. 
Some of global stellar parameters of the reference model are as follows: 
$M=1.36M_{\odot}$, $Y_{\mathrm{init}}=0.26$, $Z_{\mathrm{init}}=0.002$, $\mathrm{Age}=2.160 \times 10^{9}$ years; 
the star is represented by a relatively low-mass stellar model around the terminal age main-sequence stage, 
exhausting most of the hydrogen at the central hydrogen-burning region. 
For more information on the reference model, see \edit1{\citet{Hatta2021}}. 
 
\section{Method} \label{sec:2}

\subsection{Rotation inversion} \label{sec:2-1}

It is relatively simple to mathematically describe the \edit1{frequency} splitting \edit1{caused by rotation 
(the rotational splitting)} 
when the internal rotation is slow compared with the dynamical timescale and 
the oscillation periods of the system\edit1{.} 
\edit1{In this case, }we can treat \edit1{rotation} as a small perturbation to the system, 
and then, we can relate the internal rotation to the rotational shifts in frequencies 
based on the first-order perturbative approach 
\citep[see more details in e.g.][]{Unno_text,Aerts_text}. 
The explicit form for the rotational shift ($d_{i}=\delta \omega_{\edit1{i}}/m$, where $i$ represents 
a particular set of mode indices\edit1{, namely,} \edit1{the radial order} $n$, \edit1{the spherical degree} $l$, 
and \edit1{the azimuthal order} $m$) thus derived is as follows: 
\begin{equation}
d_{i} = \int \!\!\! \int K_{i}(x, \mu) \Omega(x,\mu)  dx d\mu + e_{i}, \label{Eq_rot_splt_4}
\end{equation}
in which the internal rotation $\Omega(x,\mu)$ is expressed as a function of a position inside the star, 
represented by the fractional radius ($x=r/R$) and the cosine of the colatitude $\theta$ ($\mu=\mathrm{cos} \, \theta$). 
The observational uncertainty for the rotational shift is given by $e_{i}$. 
The rotational splitting kernel $K_{i}(x, \mu)$ can be obtained 
by calculating the linear adiabatic oscillation of a certain reference model, and 
the explicit form can be found in e.g. \citet{Aerts_text}. 


Then, what we have to do to estimate the internal rotation $\Omega(x,\mu)$ 
is to \edit1{invert} the set of the equations (\ref{Eq_rot_splt_4}) \edit1{(rotation inversion)}, where 
the number of the equations is identical to that of observed rotational shifts. 
Techniques such as 
the Regularized Least-Squares (RLS) method \citep[e.g.][]{Tikhonov_and_Arsenin1977} and 
the Optimally Localized Averaging (OLA) method \citep{Backus_and_Gilbert1967}, both of which are well established, 
have been frequently used in helioseismology, 
which is also the case in asteroseismology. 

Note that, in asteroseismology, we do not have a large number of rotational shifts (around a few dozens at most) 
compared with the case in helioseismology (of the order of $10^{5}$), and thus, 
it is sometimes difficult to draw definitive conclusions 
based on just one method (even when it is one of the standard methods) and 
comparisons of results obtained via different methods 
can help us to better understand the inversion results, 
which is another reason why we attempt to develop a new scheme of Bayesian rotation inversion in this study. 
In other words, there is no all-round method 
which enables us to solve any inverse problems completely, i.e., 
each inversion technique provides us with the corresponding estimate 
based on a particular criterion adopted for the technique. 
Therefore, we should be cautious not to jump to seemingly satisfactory conclusions, 
which is especially the case in asteroseismology 
where the relative scarcity of observed rotational shifts easily leads to the ill-posedness of the inverse problems. 

\subsection{Basic points in Bayesian statistics} \label{sec:2-2}
In this section, we would like to give a few basic points in Bayesian statistics. 
Bayesian statistics allows us to, for example, investigate global properties of probabilities of parameters or 
compute probabilities of models, based on the latter of which 
we can conduct model comparison among possible models. 
In particular, such capability of the model comparison 
strongly motivates us to construct an inversion scheme based on Bayesian statistics, 
and the application can be found in Section \ref{sec:4}, 
with which the possibility of the fast-convective-core rotation of KIC 11145123 is tested. 
For more thorough introductions and discussions on Bayesian statistics 
in astronomical contexts, readers should refer to e.g. \citet{Gregory2005}. 
Applications in global and local helioseismology can be also found in \citet{Kashyap2021} and \cite{Jason2020}, respectively. 

%

Then, let us introduce one of the most fundamental equations 
in Bayesian statistics, the Bayes' theorem, which has the following form: 
\begin{equation}
p(\boldsymbol{\theta}|\mathbf{d})=\frac{p(\mathbf{d}|\boldsymbol{\theta})p(\boldsymbol{\theta})}{p(\mathbf{d})},  \label{Bayes_2}
\end{equation}
which can be derived based on the definition of the conditional probability. 
Datasets (obtained by observation) and parameters (to be estimated) 
are represented by $\mathbf{d}$ and $\boldsymbol{\theta}$ following the notation in \citet{Benomar2009}. 
The probabilities $p(\boldsymbol{\theta}|\mathbf{d})$, $p(\mathbf{d}|\boldsymbol{\theta})$, 
$p(\boldsymbol{\theta})$, and $p(\mathbf{d})$ should 
read the posterior probability of the parameters given the dataset, 
the probability of the dataset given the parameters 
(or, the likelihood of the parameters), 
the prior probability of the parameters, 
and the probability of the dataset marginalized by all the parameters 
(the so-called global likelihood), respectively. 

What the Bayes' theorem (\ref{Bayes_2}) indicates is actually not complex to interpret; 
though we have to begin with uninformative states (represented by the prior probability $p(\boldsymbol{\theta})$), once we conduct observations, we can 
\edit1{estimate} the probability of obtaining the resultant datasets assuming a set of parameters (represented by the likelihood of the parameters $p(\mathbf{d}|\boldsymbol{\theta})$), and finally, 
combining the prior probability and the likelihood enables us to update our understanding of the parameters (represented by the posterior probability $p(\boldsymbol{\theta}|\mathbf{d})$). 

\subsection{Formulation} \label{sec:2-3}
One of the goals in the Bayesian rotation inversion is 
to compute the posterior probability of parameters 
describing a rotational profile given the observed rotational shifts $p(\Omega(x,\mu)|\mathbf{d})$. 
To this end, we need to determine 
functional forms of the prior probability $p(\Omega(x,\mu))$ 
and the likelihood $p(\mathbf{d}|\Omega(x,\mu))$ beforehand as shown in Subsection \ref{sec:2-3-1}. 
We also describe how to carry out model comparison in the Bayesian framework in Subsection \ref{sec:2-3-2}. 
\edit1{In this subsection, we basically follow the formulations 
in \citet{Benomar2009}.}  

\subsubsection{Priors and likelihoods} \label{sec:2-3-1}
For the prior probability, 
we adopt a uniform distribution 
whose range is reasonably broad, 
as we, a priori, do not have strong constraints on the parameters to be estimated. 
Although it is generally recognized that 
the choice of functional forms of prior probabilities could affect Bayesian inferences 
\citep[e.g.][]{Benomar2009}, this seems not to be the case in our study as will be seen in Section \ref{sec:4}, 
which is another reason for adopting simple uninformative prior probabilities for the parameters. 
%
The specific range for the prior probability is later shown in 
Sections \ref{sec:3-2} and \ref{sec:4-2}. 

For the likelihood, we assume that 
an observed rotational shift $d_{i}$ is a realization 
from a Gaussian distribution whose mean is given by 
the first term on the right-hand side in expression (\ref{Eq_rot_splt_4}) 
with its standard deviation identical to that of the corresponding observational uncertainty $e_{i}$. 
Based on the assumption, the explicit form for the likelihood of the parameters 
given the rotational shift is as below: 
\begin{eqnarray}
\lefteqn{p(d_{i}|\Omega(x,\mu))}  \nonumber \\ 
&=&\frac{1}{\sqrt{2 \pi} e_{i}} \mathrm{exp} \biggl [ -\frac{1}{2} \biggl ( \frac{d_{i} - \int K_{i}(x,\mu) \Omega (x,\mu) dx d\mu}{e_{i}} \biggr )^2 \biggr ]. \nonumber \\
  \label{Bayes_3}
\end{eqnarray}

By further assuming that the observed rotational shift is statistically independent \edit1{of} each other, 
the explicit form for the likelihood of the parameters 
given the set of the observed rotational shifts is 
the product of expressions (\ref{Bayes_3}) as follow\edit1{s}: 
\begin{equation}
p(\mathbf{d}|\Omega(x,\mu))=  \prod_{i=1}^{N} p(d_{i}|\Omega(x,\mu)),   \label{Bayes_4}
\end{equation}
where the number of the observed rotational shifts is denoted \edit1{by} $N$. 

Based on the determined prior probability and 
the likelihood of the parameters given the set of the observed rotational shifts, 
we can calculate the posterior probability of the parameters 
following expression (\ref{Bayes_2}), and we can subsequently obtain estimates for the 
parameters by, for instance, choosing a set of the parameters 
for 
which the posterior probability is maximum (called Maximum A Posteriori estimation). 

Note that we have not determined an explicit functional form 
for the rotational profile $\Omega(x,\mu)$ yet, 
which is a necessary step for us to compute the likelihood 
(we have to compute the integration inside expression \ref{Bayes_3}). 
Several specific rotational profiles (and 
the corresponding results of the Bayesian rotation inversion) can be found in Sections \ref{sec:3} and \ref{sec:4}. 
It should also be noted that calculation of the posterior probability 
requires us to carry out numerical integrations via, for instance, 
the Markov Chain Monte Carlo (MCMC) method 
if the number of the parameters used to describe a rotational profile 
is so large that it is computationally impossible to directly evaluate the posterior probability. 

\subsubsection{Model comparison based on global likelihoods} \label{sec:2-3-2}
In this subsection, we would like to mention a model comparison based on Bayesian statistics. 
The important quantity is the global likelihood $p(\mathbf{d})$, 
which is 
a normalization constant in expression (\ref{Bayes_2}). 
We can confirm the importance of the global likelihood 
by reconsidering the Bayes' theorem (\ref{Bayes_2}), which can be rewritten as 
\begin{equation}
p(\boldsymbol{\theta}|\mathbf{d},M_{j})=\frac{p(\mathbf{d}|\boldsymbol{\theta},M_{j})p(\boldsymbol{\theta},M_{j})}{p(\mathbf{d}|M_{j})},  \label{Bayes_5}
\end{equation}
where a model $M_{j}$ representing a certain set of parameters is 
explicitly expressed, and the global likelihood can read the likelihood of the model $M_{j}$ given the dataset. 

Then, let us consider the posterior probability of 
the model $M_{j}$ 
as below: 
\begin{equation}
p(M_{j}|\mathbf{d})=\frac{p(\mathbf{d}|M_{j})p(M_{j})}{p(\mathbf{d})},  \label{Bayes_6}
\end{equation}
and let us compare the posterior probability of the model $M_{j}$ and that of another model $M_{k}$. 
Taking the ratio between the two posterior probabilities \citep[called odds ratio $O_{M_{j},M_{k}}$;][]{Gregory2005} leads to 
\begin{equation}
O_{M_{j},M_{k}} = \frac{p(M_{j}|\boldsymbol{d})}{p(M_{k}|\boldsymbol{d})} = \frac{p(\boldsymbol{d}|M_{j})}{p(\boldsymbol{d}|M_{k})},  \label{Bayes_7}
\end{equation}
in which the ratio of the posterior probabilities of the models 
is expressed by the ratio of the global likelihoods of the models (note that 
it is assumed that $p(M_{j})=p(M_{k})$ here). 

\edit1{It is conventionally} 
\edit2{considered} \edit1{that the model $M_{j}$ is substantially favored compared with the model $M_{k}$ 
when $O_{M_{j},M_{k}} > 3$ \citep[e.g.][]{Jeffreys1998}.} 
\edit2{This threshold can be explained with a simple example where 
there are just two models $M_{j}$ and $M_{k}$. 
In that case, the posterior probability of the model $M_{j}$ can be expressed as 
$[1 + O_{M_{j},M_{k}}^{-1}]^{-1}$. Thus, the threshold $3$ corresponds to the probability $0.75$.} 
In this way, the global likelihoods $p(\mathbf{d}|M)$ are 
such essential quantities that we can select the most probable model 
given the dataset, which is practically demonstrated in Sections \ref{sec:3} and \ref{sec:4}.

\section{Simple tests} \label{sec:3}
The Bayesian scheme explained in Section \ref{sec:2} is tested in this section. 
We firstly introduce two models of rotational profile, from which 
two specific rotational profiles, 
one with a rotational velocity shear and the other without it, are artificially generated. 
The corresponding sets of rotational shifts are given as well (Section \ref{sec:3-1}). 
We carry out Bayesian rotation inversion with the artificial datasets \edit1{to compute the posterior probabilities 
and global likelihoods, based on which} 
whether we can \edit1{choose} the correct model of 
rotational profile or not \edit1{is checked }(Section \ref{sec:3-2}). 

\subsection{Artificial rotational shifts} \label{sec:3-1}
\begin{figure} [t]
\begin{center}
\includegraphics[scale=0.30,angle=-90]{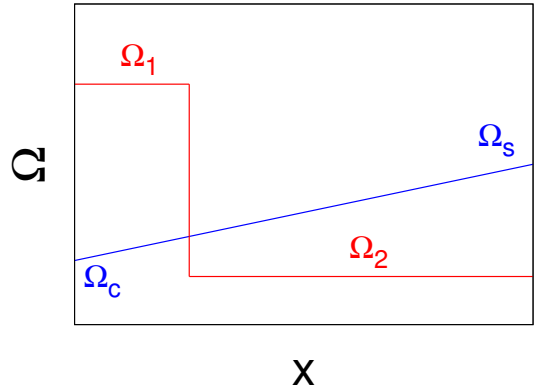}
\caption{\footnotesize Schematic view for two models of internal rotation \edit1{profile} $\Omega(x)$, namely, 
$M_{\mathrm{sh}}$ (red) and $M_{\mathrm{lin}}$ (blue), 
which are described with the sets of two parameters $(\Omega_{1},\Omega_{2})$ 
and $(\Omega_{\mathrm{c}},\Omega_{\mathrm{s}})$, respectively. 
}
\label{parameterization_rot_profs_simple_test}
\end{center} 
\end{figure}

A rotational shift of a certain mode $d_{i}$ can be computed following expression (\ref{Eq_rot_splt_4}) 
once we specify a ``true'' rotational profile 
$\Omega (x,\mu)$ and 
calculate the rotational splitting kernel $K_{i}(x,\mu)$ of the model. 
We present two models of 
rotational profile 
$\Omega (x)$ as below (note that 
we do not consider latitudinal dependence of the internal rotation for simplicity in this section). 
The first one is described with two parameters, namely, a rotation rate of the core $\Omega_{1}$ 
and that of the envelope $\Omega_{2}$ (the red lines in Figure \ref{parameterization_rot_profs_simple_test}). 
The other one has a linear profile in terms of the fractional radius, which is parameterized with 
a rotation rate of the center of the star $\Omega_{\mathrm{c}}$ and the surface rotation rate $\Omega_{\mathrm{s}}$ 
(the blue line in Figure \ref{parameterization_rot_profs_simple_test}). 
Let us call the former model 
$M_{\mathrm{sh}}$ and 
the latter one 
$M_{\mathrm{lin}}$. 
Specific parameters for two particular rotational profiles are here given as below: 
\begin{equation}
(\Omega_{1,\mathrm{true}},\Omega_{2,\mathrm{true}})=(\edit1{1.50},\edit1{0.980}),  \label{Bayes_8}
\end{equation}
and 
\begin{equation}
(\Omega_{\mathrm{c},\mathrm{true}},\Omega_{\mathrm{s},\mathrm{true}})=(0.950,0.980),  \label{Bayes_9}
\end{equation}
in units of $2 \pi \times 0.01 \, d^{-1}$. 
Let us call the rotational profiles thus specified 
$\Omega_{\mathrm{sh}}$ and 
$\Omega_{\mathrm{lin}}$, respectively. 

\edit1{Note} that, in this study, we concentrate on main-sequence stars with a convective core as is the case for KIC 11145123. 
\edit1{Accordingly}, 
we assume that the boundary in $\Omega_{\mathrm{sh}}$ is fixed to be the convective boundary; 
$\Omega_{1}$ and $\Omega_{2}$ represent  
the rotation rate of the convective core and that of the radiative envelope. 

For computing splitting kernels $K_{i}(x,\mu)$, 
the linear adiabatic oscillation of the reference model, which is the non-standard model of 
KIC 11145123 constructed by \edit1{\citet{Hatta2021}} (see Section \ref{sec:intro}), has been calculated 
via a linear adiabatic oscillation code GYRE \citep{Townsend2013}. 
Based on the eigenfunctions and eigenfrequencies thus obtained, 
we calculate the splitting kernels $K_{i}(x,\mu)$ \citep[the explicit form can be found in, e.g.,][]{Aerts_text}. 
We have computed splitting kernels for 33 eigenmodes, namely, 20 high-order g modes with $(l,m)=(1,1)$, 
3 low-order p modes with $(l,m)=(1,1)$, 5 low-order mixed modes with $(l,m)=(2,1)$, 
and 5 low-order mixed modes with $(l,m)=(2,2)$. 
The eigenmodes are thus chosen since these types of modes 
(high-order g modes and low-order p/mixed modes) are 
frequently observed for $\gamma$ Dor-$\delta$ Sct type hybrid stars such as KIC 11145123 
though the relatively larger number of modes compared with the actual observation are prepared 
to render rotation inversion as robust as possible. 
It should be noted that the results of the simple test here are qualitatively the same even if 
we 
use a smaller set of splitting kernels, which is identical to that in the case of KIC 11145123 (see Section \ref{sec:4-1}). 

Then, based on the rotational profiles, namely, 
$\Omega_{\mathrm{sh}}$ and 
$\Omega_{\mathrm{lin}}$, and the splitting kernels $K_{i}(x,\mu)$, 
we have computed the corresponding sets of rotational shifts following expression (\ref{Eq_rot_splt_4}), 
in which the observational uncertaint\edit1{ies} $e_{i}$ \edit1{are} assumed to be realization\edit1{s} from a 
Gaussian distribution whose mean and standard deviation are $0$ and $10^{-3}$ (in units of $2 \pi \times 0.01 \, d^{-1}$). 
The standard deviation of the Gaussian distribution is chosen so that it is 
around a typical observational uncertainty of the rotational shifts for KIC 11145123. 
It is also assumed that the observational uncertainty $e_{i}$ is statistically independent \edit1{of} each other. 
We thus have two sets of 33 artificially generated rotational shifts with which Bayesian rotation inversion is to be carried out in the following 
sections. 
We denote the artificial dataset for the rotational profile with a velocity shear as $\delta \omega_{\mathrm{sh}}$
and that for the linear rotational profile as $\delta \omega_{\mathrm{lin}}$. 
\begin{figure} [t]
\begin{center}
\includegraphics[scale=0.45]{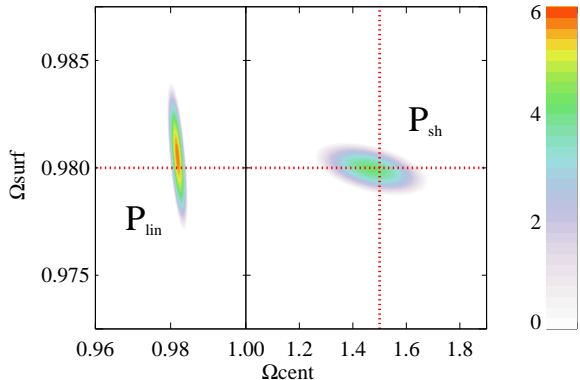}
\caption{\footnotesize \edit1{Posterior probability density function 
given the artificially generated dataset $\delta \omega_{\mathrm{sh}}$ 
computed based on $M_{\mathrm{sh}}$ (labeled by $\mathrm{P}_{\mathrm{sh}}$) 
and that computed based on $M_{\mathrm{lin}}$ (labeled by $\mathrm{P}_{\mathrm{lin}}$). 
For convenience, the common parameters $(\Omega_{\mathrm{cent}}, \Omega_{\mathrm{surf}})$ are used 
to describe the rotation rate of the center (the horizontal axis) and that of the surface (the vertical axis), 
both of which are in units of $2 \pi \times 0.01 \, d^{-1}$. 
The posterior probability densities (in units of $(100/ 2 \pi)^2 \, d^{2}$) are expressed in the decimal logarithm (see the color bar). 
The parameters prepared beforehand (\ref{Bayes_8}) are represented by the red dotted lines. 
}
 }
\label{2d_log_posts_revised_splt2z}
\end{center} 
\end{figure}

\edit1{\subsection{Bayesian rotation inversion with the artificial rotational shifts} \label{sec:3-2}} 
In this section, the Bayesian scheme has been utilized 
with the set of rotational shifts \edit1{(}$\delta \omega_{\mathrm{sh}}$ \edit1{or $\delta \omega_{\mathrm{lin}}$}\edit1{)}, 
in order to test whether we can correctly \edit1{choose} the right model 
of rotational profile ($M_{\mathrm{sh}}$ \edit1{or $M_{\mathrm{lin}}$, respectively}) 
via the Bayesian scheme or not. 

As mentioned in Subsection \ref{sec:2-3-1}, the first thing we have to do is to 
specify parameters to describe $\Omega(x,\mu)$ for 
which the posterior probabilities are to be computed. 
For simplicity, we have used the models of rotational profile 
$M_{\mathrm{sh}}$ described by $(\Omega_{1},\Omega_{2})$ 
and $M_{\mathrm{lin}}$ described by $(\Omega_{\mathrm{c}},\Omega_{\mathrm{s}})$ 
(Figure \ref{parameterization_rot_profs_simple_test}), 
the former of which \edit1{contains $\Omega_{\mathrm{sh}}$} 
and the latter of which 
\edit1{contains $\Omega_{\mathrm{lin}}$}. 
%

Secondly, prior probabilities for the prepared parameters have to be specified. 
We assume that the prior probabilities are uniform as below: 
\begin{equation}
\Omega_{1}, \, \Omega_{2}, \,  \Omega_{\mathrm{c}}, \, \Omega_{\mathrm{s}} \sim U[0,50],  \label{Simple_test_prior} 
\end{equation}
where $q\sim U[a,b]$ means that $q$ is a random variable uniformly distributed in a range from $a$ to $b$. 
The rotation rates are in units of $2 \pi \times 0.01 \, d^{-1}$. 
The joint prior probability is computed by taking products of prior probabilities assuming 
that the parameters are statistically independent \edit1{of} each other. 

Based on the priors and likelihoods, 
which can be computed with expressions (\ref{Bayes_3}) and (\ref{Bayes_4}), 
we calculate the numerator of the right hand side in expression (\ref{Bayes_2}), 
integrate it over the parameter space to obtain the normalization constant $p(\mathbf{d})$ (or, the global likelihood), 
and finally compute the posterior probability of parameters 
%
%
given the dataset. 
Since the numbers of the parameters are just two for both of the models and 
it is not computationally expensive to numerically carry out such 2-dimensional computations, 
we have directly calculated the posterior probabilities. 

\begin{figure} [t]
\begin{center}
\includegraphics[scale=0.45]{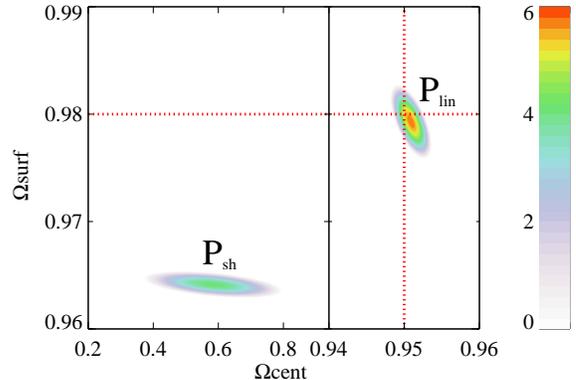}
\caption{\footnotesize \edit1{Same as Figure \ref{2d_log_posts_revised_splt2z} 
except that $\delta \omega_{\mathrm{lin}}$ is used instead of $\delta \omega_{\mathrm{sh}}$. 
The parameters prepared beforehand (\ref{Bayes_9}) are represented by the red dotted lines. }
}
\label{2d_log_posts_revised_splt1_expanded}
\end{center} 
\end{figure}
\begin{table} [t]
 \begin{center}
  \caption{\footnotesize \edit1{Odds ratios $O_{M_{i}, M_{j}}$ expressed in the decimal logarithm, 
  in the cases with $\delta \omega_{\mathrm{sh}}$ (top) 
  and $\delta \omega_{\mathrm{lin}}$ (bottom)}}
    \begin{tabular} {c|cc} \hline\hline
 \edit1{$M_{i} \backslash M_{j} $} & $M_{\mathrm{sh}}$ &  $M_{\mathrm{lin}}$  \\ \hline
\edit1{$M_{\mathrm{sh}}$} & \edit1{0}  & \edit1{23.10} \\ 
\edit1{$M_{\mathrm{lin}}$} &  \edit1{-23.10} &  \edit1{0}  \\ \hline
\addlinespace[2.5mm] 
    \end{tabular}
    \\
    \begin{tabular} {c|cc} \hline\hline
 \edit1{$M_{i} \backslash M_{j} $} & $M_{\mathrm{sh}}$ &  $M_{\mathrm{lin}}$  \\ \hline
\edit1{$M_{\mathrm{sh}}$} & \edit1{0}  & \edit1{-135.6} \\ 
\edit1{$M_{\mathrm{lin}}$} &  \edit1{135.6} &  \edit1{0}  \\ \hline
    \end{tabular}
    \\ 
\end{center}
\footnotesize \edit1{$\mathbf{Note.}$ Based on the definition (\ref{Bayes_7}), it is 
apparent that 
$O_{M_{i}, M_{j}} = (O_{M_{j}, M_{i}})^{-1}$. 
It is generally considered that the model $M_{i}$ is substantially favored compared with the model $M_{j}$ when 
$\mathrm{log} \,O_{M_{i}, M_{j}} > 0.5$.} 
 \label{Glob_like_Ch3}
\end{table}

Figure \ref{2d_log_posts_revised_splt2z} 
\edit1{shows two posterior probability density functions} \edit1{given $\delta \omega_{\mathrm{sh}}$.} 
\edit1{One is computed based on} $M_{\mathrm{sh}}$ 
and the other is \edit1{computed based on }
$M_{\mathrm{lin}}$ \edit1{($\mathrm{P}_{\mathrm{sh}}$ and 
$\mathrm{P}_{\mathrm{lin}}$, respectively, in Figure \ref{2d_log_posts_revised_splt2z}).}
\edit1{Since }both of the posterior probabilities are unimodal, we 
can estimate the parameters by, for instance, taking a parameter set that maximizes the 
corresponding posterior probability (the Maximum A Posteriori estimate). 

\edit1{It is apparent in Figure \ref{2d_log_posts_revised_splt2z} that 
the Maximum A Posteriori estimate for $\mathrm{P}_{\mathrm{sh}}$ 
is almost identical to the prepared parameters (\ref{Bayes_8}) (red dotted lines in Figure \ref{2d_log_posts_revised_splt2z}) 
while those for $\mathrm{P}_{\mathrm{lin}}$ are biased.}
The important point, however, is that we cannot determine which model describes 
the dataset $\delta \omega_{\mathrm{sh}}$ 
more appropriately by just comparing the posterior probabilities 
\edit1{without knowing the prepared parameters (\ref{Bayes_8}) beforehand.}
As it is described in Subsection \ref{sec:2-3-2}, such model comparison \edit1{should} be 
achieved by comparing the global likelihood, which is the normalization constant 
in expression (\ref{Bayes_2}). 

In \edit1{the case of }this simple test with $\delta \omega_{\mathrm{sh}}$, we have obtained the following 
\edit1{odds ratio (defined by equation (\ref{Bayes_7})): $\mathrm{log}\, O_{M_{\mathrm{sh}},M_{\mathrm{lin}}} = 23.10$} 
(Table \ref{Glob_like_Ch3})\edit1{, 
which is much larger than the conventionally adopted criterion 
$\mathrm{log}\, 3 \sim 0.5$ \citep[e.g.][]{Jeffreys1998}.} 
It is therefore correctly inferred that, in the light of the Bayesian scheme, the model of rotational profile with a velocity shear 
$M_{\mathrm{sh}}$ 
is more favorable to describe the artificially generated dataset $\delta \omega_{\mathrm{sh}}$. 

\edit1{The same is true} \edit2{when} 
\edit1{$\delta \omega_{\mathrm{lin}}$ is used for the test 
(Figure \ref{2d_log_posts_revised_splt1_expanded}). 
The obtained 
\edit1{odds ratio is $\mathrm{log}\, O_{M_{\mathrm{lin}},M_{\mathrm{sh}}} = 135.6$} 
(Table \ref{Glob_like_Ch3}). 
It is} \edit2{clear} 
\edit1{that the model $M_{\mathrm{lin}}$ is preferred, 
which is the right one we have used to generate the dataset $\delta \omega_{\mathrm{lin}}$. 
It should also be instructive to mention that the estimates are biased unless we have chosen the correct model 
(see $\mathrm{P}_{\mathrm{sh}}$ in Figure \ref{2d_log_posts_revised_splt1_expanded}), 
highlighting the importance of the model comparison in the Bayesian context achieved by computing global likelihoods. }

\section{Applying the method to KIC 11145123} \label{sec:4}
The Bayesian scheme demonstrated in the previous sections is applied to one of the Kepler targets, KIC 11145123, 
to infer its internal rotation profile, especially focusing on the convective-core rotation. 
After we present the rotational shifts and corresponding splitting kernels for KIC 11145123 (Section \ref{sec:4-1}), 
basic setups for 
the Bayesian rotation inversion are given (Section \ref{sec:4-2}), 
based on which the posterior probabilities and the global likelihoods are computed, and the model 
comparison is conducted as well (Section \ref{sec:4-3}). 
\edit1{Then, the validation of the results obtained is carried out in Section \ref{sec:4-val}.} 
We \edit1{finally} provide a brief discussion on the results in Section \ref{sec:4-4}. 

\subsection{Data and splitting kernels} \label{sec:4-1}
We have used a set of rotational shifts and observational uncertainties 
which have been measured and determined by \citet{Kurtz2014} (see 
Tables 1 and 2 in the paper). 
The set is composed of twenty-three eigenmodes, namely, 
fifteen high-order g modes with $(l,m)=(1,1)$, two low-order p modes with $(l,m)=(1,1)$, 
three low-order mixed modes with $(l,m)=(2,1)$, and three low-order mixed modes with $(l,m)=(2,2)$. 
The mode identification for the mode set has been conducted based on the non-standard model of the star 
constructed by \edit1{\citet{Hatta2021}}, 
according to which the star is a low-mass star at the terminal age main-sequence stage. 
Some of basic global parameters of the non-standard model can also be found in the last paragraph of Section \ref{sec:intro}. 

\begin{figure} [t]
\begin{center}
\includegraphics[scale=0.45]{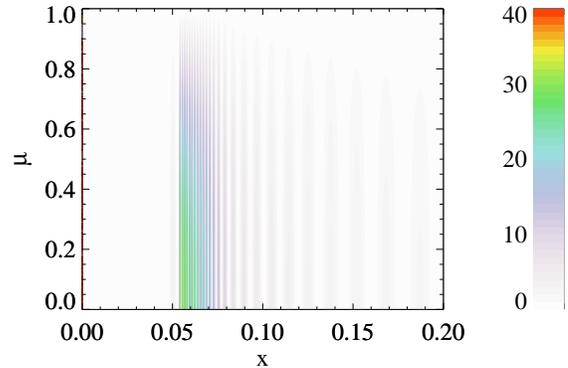}
\caption{\footnotesize \edit1{Splitting kernel of a high-order g mode with $(n,l,m)=(-34,1,1)$ for KIC 11145123. 
The horizontal and vertical axes are the fractional radius $x$ and the cosine of the colatitude $\mu$, respectively. 
}
}
\label{g_kernel_tentative_expand}
\end{center} 
\end{figure}
\begin{figure} [t]
\begin{center}
\includegraphics[scale=0.45]{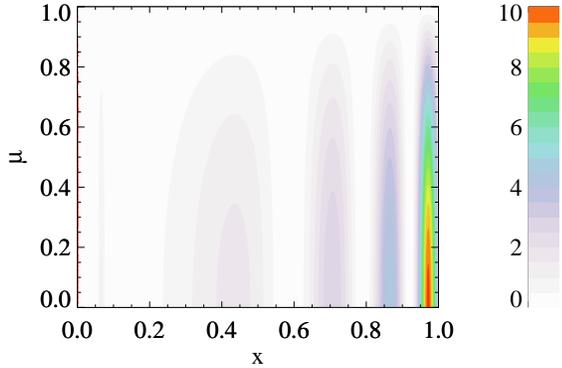}
\caption{\footnotesize \edit1{Same as Figure \ref{g_kernel_tentative_expand}, but for a low-order p mode with $(n,l,m)=(3,1,1)$. 
}
}
\label{p1_kernel_l1n3_tentative}
\end{center} 
\end{figure}
\begin{figure} [t]
\begin{center}
\includegraphics[scale=0.45]{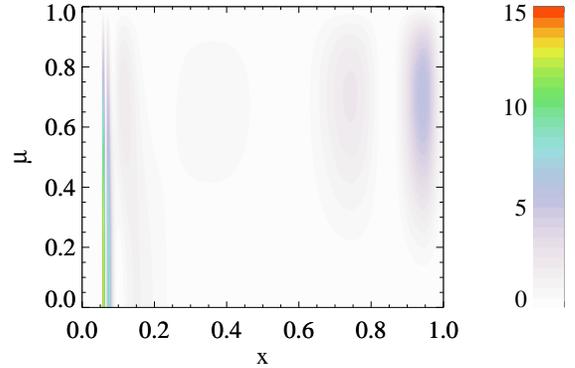}
\caption{\footnotesize \edit1{Same as Figure \ref{g_kernel_tentative_expand}, but for a mixed mode with $(n,l,m)=(-1,2,1)$. 
}
}
\label{p2_kernel_l2nm1m1_tentative}
\end{center} 
\end{figure}

The corresponding splitting kernels are calculated using the non-standard model as a reference model. 
We have 
three types of the splitting kernels. 
The first type corresponds to high-order g modes and it has sensitivity in the deep radiative region just above the convective core 
(Figure \ref{g_kernel_tentative_expand}). 
The second type and the third correspond to low-order p modes and low-order mixed modes, respectively 
(Figures \ref{p1_kernel_l1n3_tentative} and \ref{p2_kernel_l2nm1m1_tentative}). 
Both of them have sensitivity in the outer envelope, but the low-order mixed modes with $(l,m)=(2,1)$ have 
sensitivity in the high-latitude region as well, which is not the case for other modes with $(l,m)=(1,1)$ or $(l,m)=(2,2)$. 
\edit1{It should be} emphasize\edit1{d} that only the mixed-mode splitting kernels 
have finite sensitivity inside the convective core
, which renders the detection of the convective-core rotation possible as it has been 
shown in \edit1{\citet{Hatta2019}.} 
A close look into the splitting kernels around the convective boundary 
is later presented and discussed in Section \ref{sec:4-4}. 


\subsection{Models of rotational profile and priors} \label{sec:4-2}
\begin{figure} [t]
\begin{center}
\includegraphics[scale=0.30,angle=-90]{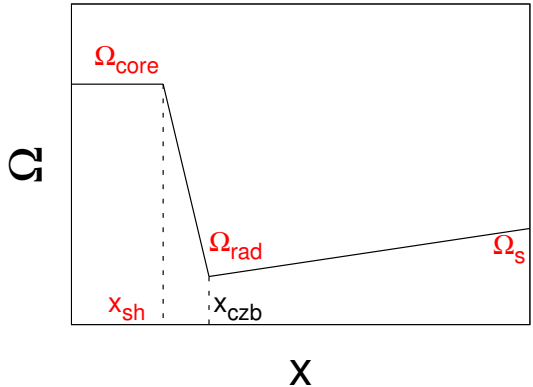}
\caption{\footnotesize Schematic picture for a way of parameterizing 
the function $\Omega_{\mathrm{0}}(x)$ in expression (\ref{Mod_comp_1}), 
which is used for the models $M_{\mathrm{1d4p}}$ and $M_{\mathrm{2d6p}}$. 
The profile contains a velocity shear, and it is described by four parameters, 
namely, the rotation rate inside the convective core $\Omega_{\mathrm{core}}$, 
that at the outer boundary of the velocity shear $\Omega_{\mathrm{rad}}$, 
that at the surface $\Omega_{\mathrm{s}}$, and the position of one of the 
boundaries of the velocity shear $x_{\mathrm{sh}}$. 
Note that another boundary of the velocity shear is fixed to be the convective boundary 
(represented as $x_{\mathrm{czb}}$ in this figure); 
$x_{\mathrm{sh}}$ could be either the inside or the outside of the convective core. 
Except for the inner core which is assumed to be rotating rigidly with $\Omega_{\mathrm{core}}$, 
rotation rates of the other regions are linearly expressed. }
\label{parameterization_rot_prof_Ch4}
\end{center} 
\end{figure}

For carrying out Bayesian rotation inversion in the case of KIC 11145123, 
%
we parameterize the rotational profile as below: 
\begin{equation}
\Omega(x,\mu) = \Omega_{0}(x) + \mu^2 \Omega_{1}(x), \label{Mod_comp_1} 
\end{equation}
where $\Omega_{0}(x)$ can have a linear profile 
described with two parameters, namely, the rotation rate at the center $\Omega_{\mathrm{c}}$ and 
that at the surface $\Omega_{\mathrm{s}}$ 
(see the blue line in Figure \ref{parameterization_rot_profs_simple_test}) 
or can have a velocity shear (with four parameters, namely, 
the rotation rate below a velocity shear boundary $x_{\mathrm{sh}}$ 
which is assumed to be uniform at $\Omega_{\mathrm{core}}$, 
the rotation rate $\Omega_{\mathrm{rad}}$ at another side of the shear boundary 
fixed to be the convective core boundary ($x_{\mathrm{czb}} \sim 0.045$), 
and that at the surface $\Omega_{\mathrm{s}}$) 
(see Figure \ref{parameterization_rot_prof_Ch4}). 

Another function $\Omega_{1}(x)$, which is related to the latitudinal dependence of the internal rotation, 
can be zero everywhere 
(with no parameters) or can have a linear profile 
(with two parameters, in almost the same way as a linear profile of $\Omega_{0}(x)$ 
but with additional indices as $\Omega_{1\mathrm{c}}$ and $\Omega_{1\mathrm{s}}$). 
We have taken such latitudinal dependence of the internal rotation into account in order to 
evaluate the effect on the global likelihood though it is not a primary subject to be investigated in this study. 
We note that the final inference on the convective-core rotation is not qualitatively changed 
due to the inclusion of the latitudinal dependence of the rotation in the analysis as we see later in Section \ref{sec:4-3}. 

With the definitions for $\Omega_{0}(x)$ and $\Omega_{1}(x)$, there are four ways of parameterization of the rotational profile in total. 
Let us denote the models as follows: $M_{\mathrm{1d2p}}$ for linear $\Omega_{0}$ and zero $\Omega_{1}$, 
$M_{\mathrm{1d4p}}$ for shear $\Omega_{0}$ and zero $\Omega_{1}$, 
$M_{\mathrm{2d4p}}$ for linear $\Omega_{0}$ and linear $\Omega_{1}$, 
and $M_{\mathrm{2d6p}}$ for shear $\Omega_{0}$ and linear $\Omega_{1}$, 
where the subscripts $\mathrm{d}$ and $\mathrm{p}$ stand for the dimension of a modeled rotational profile 
and the number of parameters in the model, respectively (Table \ref{Glob_like_Ch4}). 

Then, what we have to specify is the corresponding priors for the parameters. 
As described in Section \ref{sec:3-2}, each prior probability for a certain parameter is assumed to be uniform as below: 
\begin{eqnarray}
\Omega_{\mathrm{c}}, \, \Omega_{\mathrm{s}}, \, \Omega_{\mathrm{core}}, \, \Omega_{\mathrm{rad}} \sim U[0.1,30], 
\label{Mod_comp_2} 
\end{eqnarray}
\begin{equation}
\Omega_{\mathrm{1c}}, \, \Omega_{\mathrm{1s}} \sim U[-10,10], \label{Mod_comp_3} 
\end{equation}
and 
\begin{equation}
x_{\mathrm{sh}} \sim U[0.010,0.055]. \label{Mod_comp_3} 
\end{equation}
The parameters representing rotation rates are in units of $2 \pi \times 0.01 \, d^{-1}$, and 
$q\sim U[a,b]$ means that $q$ is a random variable uniformly distributed in a range from $a$ to $b$, 
as described in Section \ref{sec:3-2}. 
The joint prior probability is computed by taking products of the prior probabilities 
based on the assumption that the parameters are \edit1{statistically} independent \edit1{of} each other. 

\subsection{Results} \label{sec:4-3}
\begin{figure} [t]
\begin{center}
\includegraphics[scale=0.30,angle=-90]{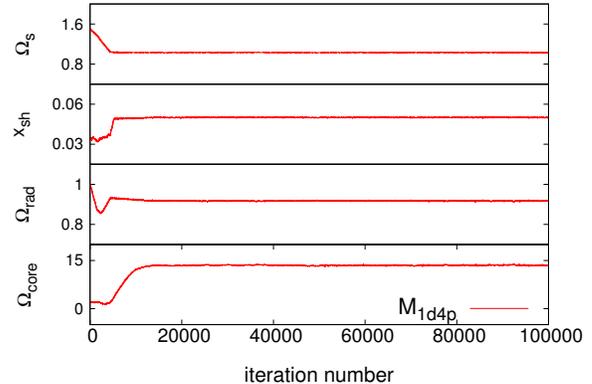}
\caption{\footnotesize Samples obtained by the Metropolis method \citep{Metropolis1953}\edit1{, 
for $\Omega_{\mathrm{s}}$ (top), $x_{\mathrm{sh}}$ (upper middle), $\Omega_{\mathrm{rad}}$ (lower middle), 
and $\Omega_{\mathrm{core}}$ (bottom)}. 
Here is the case of the \edit1{model} $M_{\mathrm{1d4p}}$. 
\edit1{The horizontal axis shows the iteration number in the sampling. }
}
\label{convergence_check_Ch4_M1d4p}
\end{center} 
\end{figure}

For each way of parameterization, the posterior probability of the parameters is computed 
based on the likelihood of the parameters and the joint prior probability. 
The likelihood is calculated with expressions (\ref{Bayes_3}) and (\ref{Bayes_4}). 
We carry out the so-called Metropolis method \citep{Metropolis1953}, 
which is one of the standard algorithms to carry out MCMC, to evaluate the 
posterior probability. 
The convergence of samples generated via the Metropolis method 
has been checked by visual inspection (see Figure \ref{convergence_check_Ch4_M1d4p}) 
with several different sets of initial values for the samples. 
A typical number of iterations required is of the order of $10^{5}$, 
and small fractions of samples are discarded from the final samples 
as they are considered as samples in the burn-in periods 
(a period during which obtained samples are not thought to be realizations 
from the posterior probability distribution we would like to sample). 
More information on principles of MCMC, how to manage the outcomes of MCMC, and so on, can be found, 
for example, in \citet{Gregory2005}. 

We consider that the posterior probabilities have been successfully sampled via MCMC, 
based on the convergence of MCMC samples 
(see Figure \ref{convergence_check_Ch4_M1d4p})\edit1{.} 
\edit1{Then, }we compute the global likelihood for each way of parameterization of the rotational profile 
to carry out the Bayesian model comparison. 
\edit1{Note that} 
\edit2{the global likelihood $p(\mathbf{d})$, which is the normalization constant 
in relation (\ref{Bayes_2}), 
cannot be determined with the MCMC samples alone.} 
\edit1{We therefore need some tools 
to compute the global likelihoods.} 
\edit1{This can be} accomplished \edit1{by} following the procedures 
proposed by \citet{Chib2001} in which an exact value of the posterior probability of a particular set of the parameters 
is directly evaluated, and then, the global likelihood is calculated based on relation (\ref{Bayes_2}). 
We have confirmed that the method correctly works for simple cases 
where we can analytically compute the posterior probability and the global likelihood. 
\begin{table} [t]
 \begin{center}
  \caption{\footnotesize \edit1{Properties of the prepared models 
 which are described by two factors, namely, the dimension of the modeled rotational profile (Dimension) and 
 the kind of the radial rotational profile of the model (Radial profile) (the top table). 
 The middle and bottom tables show odds ratios $O_{M_{i}, M_{j}}$ (expressed in the decimal logarithm) 
 computed given the observed rotational shifts with 
 or without the mixed modes, respectively. }
  }
    \begin{tabular} {c|cccc} \hline\hline
\footnotesize{Model Name}  & $M_{\mathrm{1d2p}}$ &  $M_{\mathrm{2d4p}}$ & $M_{\mathrm{1d4p}}$ &  $M_{\mathrm{2d6p}}$  \\ \hline
\footnotesize{Dimension}  & \footnotesize{1d} &  \footnotesize{2d} & \footnotesize{1d} &  \footnotesize{2d}  \\
\footnotesize{Radial profile}   & \footnotesize{Linear} &  \footnotesize{Linear} & \footnotesize{Shear} &  \footnotesize{Shear} \\ \hline\hline
  \addlinespace[2.5mm] 
      \end{tabular}
    \\ 
      \begin{tabular} {c|cccc} \hline\hline
 \edit1{$M_{i} \backslash M_{j} $} & $M_{\mathrm{1d2p}}$ &  $M_{\mathrm{2d4p}}$  &  $M_{\mathrm{1d4p}}$ & $M_{\mathrm{2d6p}}$ \\ \hline
\edit1{$M_{\mathrm{1d2p}}$} & \edit1{0}  & \edit1{-620.0}  &  \edit1{-4710} & \edit1{-5390} \\ 
\edit1{$M_{\mathrm{2d4p}}$} &  \edit1{620.0} &  \edit1{0}  &  \edit1{-4090}  & \edit1{-4770} \\ 
\edit1{$M_{\mathrm{1d4p}}$} &  \edit1{4710} &   \edit1{4090} &  \edit1{0}  & \edit1{-680.0} \\ 
\edit1{$M_{\mathrm{2d6p}}$} &  \edit1{5390} &  \edit1{4770}  &  \edit1{680.0}  & \edit1{0} \\  \hline
\addlinespace[2.5mm] 
    \end{tabular}
    \begin{tabular} {c|cc} \hline\hline
 \edit1{$M_{i} \backslash M_{j} $} & $M_{\mathrm{1d2p}}$ &  $M_{\mathrm{1d4p}}$  \\ \hline
\edit1{$M_{\mathrm{1d2p}}$} & \edit1{0}  & \edit1{-20.00} \\ 
\edit1{$M_{\mathrm{1d4p}}$} &  \edit1{20.00} &  \edit1{0}  \\ \hline
    \end{tabular}
 	\\  
\end{center}
 \footnotesize \edit1{$\mathbf{Note.}$ Based on the definition (\ref{Bayes_7}), it is 
  apparent that 
  $O_{M_{i}, M_{j}} = (O_{M_{j}, M_{i}})^{-1}$. 
  It is generally considered that the model $M_{i}$ is substantially favored compared with the model $M_{j}$ when 
$\mathrm{log} \, O_{M_{i}, M_{j}} > 0.5$.}
 \label{Glob_like_Ch4}
\end{table}

The decimal logarithm of the \edit1{odds ratios 
thus obtained are 
$\mathrm{log}\, O_{M_{\mathrm{1d4p}},M_{\mathrm{1d2p}}} = 4710$, 
$\mathrm{log}\, O_{M_{\mathrm{1d4p}},M_{\mathrm{2d4p}}} = 4090$, 
$\mathrm{log}\, O_{M_{\mathrm{2d6p}},M_{\mathrm{1d2p}}} = 5390$, 
and $\mathrm{log}\, O_{M_{\mathrm{2d6p}},M_{\mathrm{2d4p}}} = 4770$} 
(Table \ref{Glob_like_Ch4}), 
clearly indicating that the rotational profiles with the velocity shear \edit1{($M_{\mathrm{1d4p}}$ and $M_{\mathrm{2d6p}}$)} 
are more favored than those without the shear \edit1{($M_{\mathrm{1d2p}}$ and $M_{\mathrm{2d4p}}$)}. 
\edit1{T}he validity of the computation of the global likelihoods 
has been checked with several different settings in MCMC, and the 
global likelihoods have been rounded to no more than four significant figures (Table \ref{Glob_like_Ch4}). 

\begin{figure} [t]
\begin{center}
\includegraphics[scale=0.48]{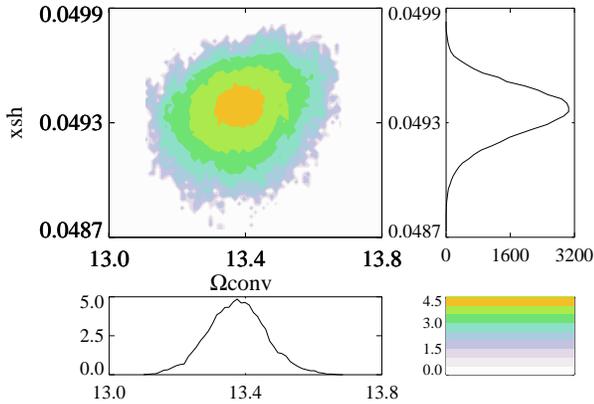}
\caption{\footnotesize \edit1{Posterior probability density function given the observed rotational shifts 
in the case that the model $M_{\mathrm{2d6p}}$ is used (top left). 
The horizontal and vertical axes represent the latitudinally averaged rotation rate of the convective core $\Omega_{\mathrm{conv}}$ 
and the position of the velocity shear $x_{\mathrm{sh}}$, in units of $2 \pi \times 0.01 \, d^{-1}$ and 
the fractional radius, respectively. 
The posterior probability density (in units of $(100/ 2 \pi) \, d\, R_{\ast}^{-1}$, where $R_{\ast}$ is the radius of the reference model) 
is expressed in the decimal logarithm (see the lower-right color bar). 
The lower-left and upper-right panels show marginalized probability density functions 
for $\Omega_{\mathrm{conv}}$ and $x_{\mathrm{sh}}$, respectively. 
}}
\label{2d_contour_map_Ch4_M2d6p_revision}
\end{center} 
\end{figure}
We can also confirm that based on the posterior probability 
for the models with the velocity shear, 
the fast-core rotation is again inferred. 
Figure \ref{2d_contour_map_Ch4_M2d6p_revision} shows the 2-dimensional contour map of 
the (marginalized) posterior probability density function 
as a function of $(\Omega_{\mathrm{\edit1{conv}}},x_{\mathrm{sh}})$, which is 
computed for the model $M_{\mathrm{\edit1{2d6p}}}$. 
\edit1{Note that $\Omega_{\mathrm{conv}}$ is used instead of $\Omega_{\mathrm{core}}$ 
to describe the latitudinally averaged rotation rate of the convective core.} 
When we adopt the \edit1{Maximum A Posteriori} estimates for the parameters, 
they are $(\Omega_{\mathrm{\edit1{conv}}},x_{\mathrm{sh}}) \sim (\edit1{13.4},\edit1{0.0493})$, 
and thus, the fast-convective-core rotation has been \edit1{inferred} as in the case of \citet{Hatta2019} 
(see Figure \ref{convergence_check_Ch4_M1d4p} for $\Omega_{\mathrm{s}}$ and $\Omega_{\mathrm{rad}}$, 
which are of the order of $1$ in units of $2 \pi \times 0.01 \, d^{-1}$ and are much slower than $\Omega_{\mathrm{\edit1{conv}}}$). 

\edit1{\subsection{Validity of the results} \label{sec:4-val}}
\edit2{Since the models used for Bayesian model comparison in the last section 
are much more complex than those in Section \ref{sec:3}, 
we would like to check whether we can conduct Bayesian model comparison even with such complex models, 
which eventually enables us to validate the results obtained in the last section. 
To this end,} 
\edit1{we present the same test as that described in Section \ref{sec:3} except for the following two points. 
Firstly, the Maximum A Posteriori estimates for the parameters, 
which are determined with the posterior probability given the model $M_{\mathrm{2d6p}}$ 
(computed in the last section), 
are used for constructing an artificial rotational profile and the 
corresponding rotational shifts. Let us call the artificial rotational shifts $\delta \omega_{\mathrm{2d6p}}$. 
Secondly, to compute the posterior probabilities given the artificial rotational shifts, 
we have used the four models, namely, $M_{\mathrm{1d2p}}$, $M_{\mathrm{1d4p}}$, 
$M_{\mathrm{2d4p}}$, and $M_{\mathrm{2d6p}}$ (see the definitions of the models in Section \ref{sec:4-2}).} 
\edit1{What we would like to check is whether we can correctly choose the right model ($M_{\mathrm{2d6p}}$ in this case) 
among the models based on the global likelihoods or not.}
\begin{figure} [t]
\begin{center}
\includegraphics[scale=0.45]{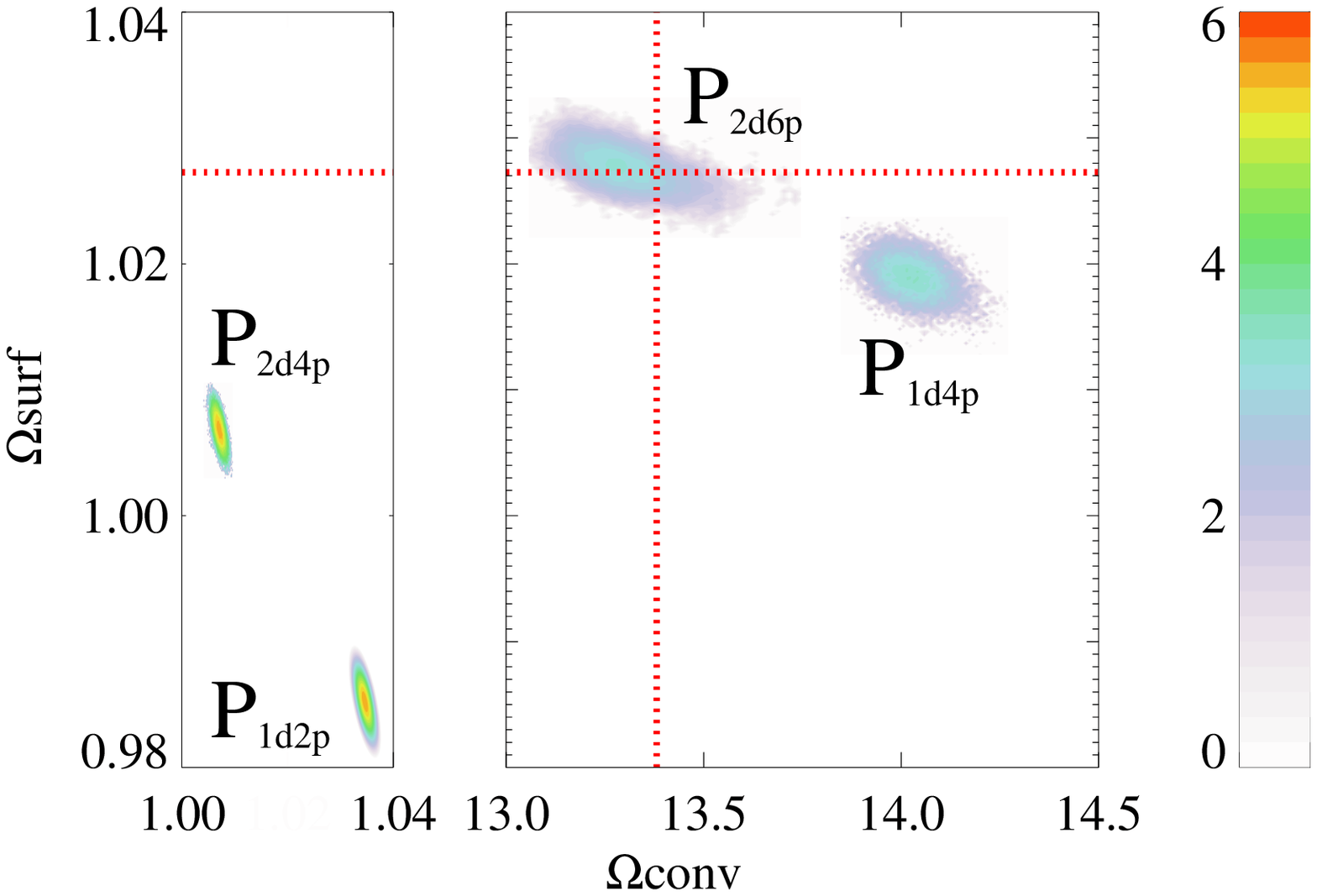}
\caption{\footnotesize \edit1{Posterior probability density functions 
given the artificially generated dataset $\delta \omega_{\mathrm{2d6p}}$ 
computed based on $M_{\mathrm{1d2p}}$ (labeled by $\mathrm{P}_{\mathrm{1d2p}}$), 
$M_{\mathrm{2d4p}}$ (labeled by $\mathrm{P}_{\mathrm{2d4p}}$), 
$M_{\mathrm{1d4p}}$ (labeled by $\mathrm{P}_{\mathrm{1d4p}}$), and 
$M_{\mathrm{2d6p}}$ (labeled by $\mathrm{P}_{\mathrm{2d6p}}$). 
For convenience, the common parameters $(\Omega_{\mathrm{conv}}, \Omega_{\mathrm{surf}})$ are used 
to describe the latitudinally averaged rotation rate of the convective core (the horizontal axis) and 
that of the surface (the vertical axis), 
both of which are in units of $2 \pi \times 0.01 \, d^{-1}$. 
The posterior probability densities (in units of $(100/ 2 \pi)^2 \, d^{2}$) 
are expressed in the decimal logarithm (see the color bar). 
The parameters prepared beforehand (which have been determined based on the model $M_{\mathrm{2d6p}}$) 
are represented by the red dotted lines. } 
}
\label{2d_log_posts_revised_astr3}
\end{center} 
\end{figure}

\edit1{The global likelihoods have been computed via the method of \citet{Chib2001}. 
The decimal logarithm of the resultant odds ratios 
are as follows: 
$\mathrm{log}\, O_{M_{\mathrm{2d6p}},M_{\mathrm{1d2p}}} = 20540$, 
$\mathrm{log}\, O_{M_{\mathrm{2d6p}},M_{\mathrm{2d4p}}} = 14390$, 
and $\mathrm{log}\, O_{M_{\mathrm{2d6p}},M_{\mathrm{1d4p}}} = 668.4$ 
(Table \ref{Glob_like_Ch5}). 
It is therefore evident that the correct model $M_{\mathrm{2d6p}}$ is preferred to the other models 
from the Bayesian perspective. 
This can be also confirmed when we see Figure \ref{2d_log_posts_revised_astr3}, where 
the prepared parameters (represented by red dotted lines in the figure) can be estimated with little 
biases only if we choose the correct model $M_{\mathrm{2d6p}}$ 
(see $\mathrm{P}_{\mathrm{2d6p}}$ in the figure). }
\begin{table} [t]
 \begin{center}
  \caption{\footnotesize \edit1{Odds ratios $O_{M_{i}, M_{j}}$ (expressed in the decimal logarithm) 
 computed given the artificially generated rotational shifts $\delta \omega_{\mathrm{2d6p}}$. }
  }
      \begin{tabular} {c|cccc} \hline\hline
 \edit1{$M_{i} \backslash M_{j} $} & $M_{\mathrm{1d2p}}$ &  $M_{\mathrm{2d4p}}$  &  $M_{\mathrm{1d4p}}$ & $M_{\mathrm{2d6p}}$ \\ \hline
\edit1{$M_{\mathrm{1d2p}}$} & \edit1{0}  & \edit1{-6150}  &  \edit1{-19870} & \edit1{-20540} \\ 
\edit1{$M_{\mathrm{2d4p}}$} &  \edit1{6150} &  \edit1{0}  &  \edit1{-13720}  & \edit1{-14390} \\ 
\edit1{$M_{\mathrm{1d4p}}$} &  \edit1{19870} &   \edit1{13720} &  \edit1{0}  & \edit1{-668.4} \\ 
\edit1{$M_{\mathrm{2d6p}}$} &  \edit1{20540} &  \edit1{14390}  &  \edit1{668.4}  & \edit1{0} \\  \hline
    \end{tabular}
 	\\  
\end{center}
 \footnotesize \edit1{$\mathbf{Note.}$ Based on the definition (\ref{Bayes_7}), it is 
  apparent that 
  $O_{M_{i}, M_{j}} = (O_{M_{j}, M_{i}})^{-1}$. 
  It is generally considered that the model $M_{i}$ is substantially favored compared with the model $M_{j}$ when 
$\mathrm{log} \, O_{M_{i}, M_{j}} > 0.5$.}
 \label{Glob_like_Ch5}
\end{table}

\subsection{Discussion} \label{sec:4-4}
Based on the results obtained in \edit1{Section \ref{sec:4-3}}, it has been \edit1{inferred}, via the Bayesian scheme, 
that the convective core of KIC 11145123 
is rotating approximately $\sim 10$ times faster than the other regions of the star. 
\edit1{We have also confirmed the validity of the results in Section \ref{sec:4-val}.} 
In this section, we have a brief discussion about the inferred position of the velocity shear. 

\begin{figure} [t]
\begin{center}
\includegraphics[scale=0.32,angle=-90]{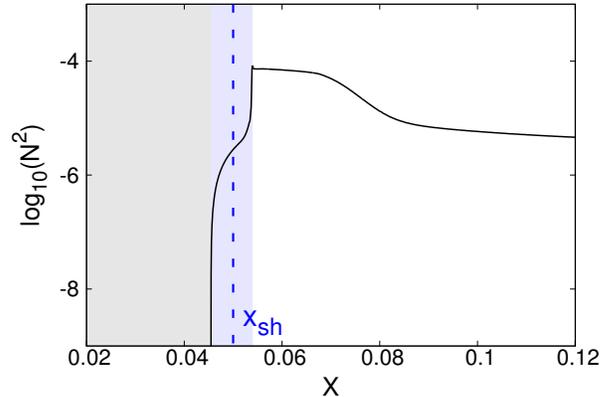}
\caption{\footnotesize Brunt-$\rm{V}\ddot{a}is\ddot{a}l\ddot{a}$ frequency \edit1{($N$)} of the reference model 
(the black curve)\edit1{ in units of $d^{-1}$.} 
\edit1{The horizontal axis shows the fractional radius $x$.} 
\edit1{T}he position of the velocity shear inferred based on the 
Bayesian rotation inversion for KIC 11145123 \edit1{is indicated with the blue dashed line.} 
\edit1{T}he deep interior is divided into three regions: the convective core (\edit1{the} grey \edit1{shaded} area), 
the overshoot zone (\edit1{the} blue \edit1{shaded} area), and the radiative region (the other area). 
}
\label{position_of_xsh}
\end{center} 
\end{figure}

\begin{figure} [t]
\begin{center}
\includegraphics[scale=0.32,angle=-90]{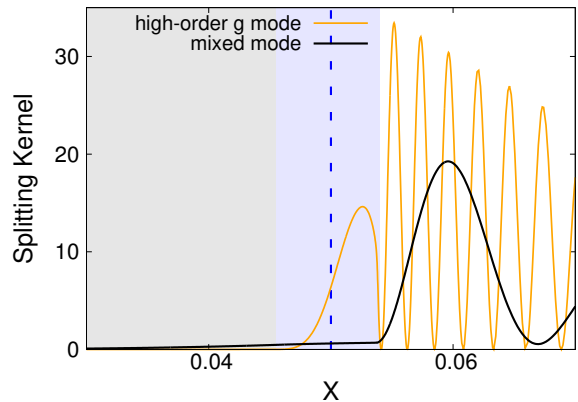}
\caption{\footnotesize Slices of the mixed-mode (black) and high-order g-mode (orange) splitting kernel\edit1{s} 
\edit1{at the equator ($\mu = 0$)}. 
\edit1{The horizontal axis shows the fractional radius $x$.} 
The inferred position of the velocity shear is represented by the blue dashed line. 
\edit1{The convective core and the overshoot zone are represented by 
the grey shaded area and the blue shaded area, respectively.}
}
\label{mixed_vs_g_modes_slice_of_kernel}
\end{center} 
\end{figure}

\edit1{Figure \ref{position_of_xsh} shows 
the Brunt-$\rm{V}\ddot{a}is\ddot{a}l\ddot{a}$ frequency of the \edit1{reference} model (black curve) 
and the inferred position of the velocity shear (blue dashed line).} 
What we would like to point out is that the inferred position 
of the velocity shear $x_{\mathrm{sh}} \sim 0.05$ is slightly above the convective boundary $x_{\mathrm{czb}}\sim 0.045$\edit1{,} 
\edit1{at which the square of the Brunt-$\rm{V}\ddot{a}is\ddot{a}l\ddot{a}$ frequency is zero, and that} 
the shear is located in the overshoot zone 
(represented as \edit1{the blue} shaded area in Figure \ref{position_of_xsh}). 
\edit1{In the overshoot zone, where overhooting has been modeled 
as a diffusive process following \citet{Herwig2000} in the case of the reference model, 
the square of the Brunt-$\rm{V}\ddot{a}is\ddot{a}l\ddot{a}$ frequency is positive 
so that the overshoot zone is included in the g-mode cavity.} 
\edit1{This} is fairly relevant to the inversion analysis since not only the mixed-modes 
but also the high-order g modes should also have 
sensitivity for the fast-core rotation (compare the black curve and 
the orange curve in Figure \ref{mixed_vs_g_modes_slice_of_kernel}). 

The relevance described in the last paragraph can be actually confirmed by a test, 
in which the Bayesian rotation inversion is carried out 
without the mixed-mode rotational shifts. 
We then have the following \edit1{odds ratio}
in the decimal logarithmic scale: 
\edit1{$\mathrm{log}\, O_{M_{\mathrm{1d4p}},M_{\mathrm{1d2p}}} = 20$}, 
(Table \ref{Glob_like_Ch4}). 
Note that the latitudinal dependence is not considered here for simplicity. 
Although the difference in the global likelihoods are much smaller 
than those computed with the mixed modes included in the analysis, 
we can still claim that the model with a velocity shear $M_{\mathrm{1d4p}}$ 
is more favorable in terms of the global likelihood; the inference of the velocity shear has just become more marginal. 

\edit1{The same analysis with the artificial rotational shifts $\delta \omega_{\mathrm{sh}}$ in Section \ref{sec:3-2} has shown} 
that \edit1{it becomes rather marginal to} 
distinguish the two models of rotational profile (one with a velocity shear \edit1{($M_{\mathrm{sh}}$)} 
and the other without it \edit1{($M_{\mathrm{lin}}$ )}) if we exclude the mixed modes\edit1{. 
The odds ratio expressed in the decimal logarithm 
is $\mathrm{log}\, O_{M_{\mathrm{sh}},M_{\mathrm{lin}}} = 1.25$.} 
\edit1{This is mainly} because we artificially locate the position of the shear identical to the convective core boundary and there are 
no modes which are sensitive to the artificially generated fast-convective-core rotation other than the mixed modes. 
However, in the case of KIC 11145123, high-order g modes can be sensitive to the fast-core rotation which slightly 
invade the radiative region, possibly leading to the marginal preference to 
the model with a velocity shear $M_{\mathrm{1d4p}}$ despite the absence of the mixed-mode rotational shifts in the analysis. 


\section{Summary} \label{sec:5}
With the increasing number of stars for which asteroseismic rotation inversion can be carried out, 
developing and improving inversion techniques is of great importance for us to 
render our inferences on the internal rotation of stars as robust and reliable as possible. 
In this study, we first present a scheme of Bayesian rotation inversion which enables us to compute the 
\edit1{probability} of a model of rotational profile and thus select the most reliable model 
among multiple models prepared by us beforehand (model comparison via the global likelihood). 
\edit1{We then conduct a simple test for the scheme using} two models of rotational profile, 
based on which two specific sets of rotational profiles and the corresponding rotational shifts are artificially generated. 
It has been shown that we can successfully \edit1{choose} the correct model of rotational profile \edit1{among the prepared models}. 

Then, we have applied the Bayesian scheme to one of the Kepler targets, KIC 11145123, for which 
the fast-convective-core rotation has been suggested by \citet{Hatta2019}. 
Focusing on the convective-core rotation of the star, four models of rotational profile are constructed. 
The global likelihoods of the four models thus computed clearly indicate that the models with the fast-convective-core rotation is 
favored, supporting the previous suggestion by \citet{Hatta2019} from a Bayesian perspective. 
\edit1{The estimated parameters have been used to construct an artificial rotational profile and the corresponding rotational shifts, 
based on which the validity of the obtained results have been checked.} 

In addition to the inference on the convective-core rotation, 
it has been suggested that the position of the rotational velocity shear is not \edit1{at} the convective boundary 
but located within the overshoot zone. 
Since it is generally considered that there are still numerous uncertainties in physics around the boundary 
between the convective core and the radiative region above for early-type main-sequence stars 
(e.g., the position of the convective boundary, the extent of the overshoot zone, the possible rotational velocity shear, 
the possible dynamo mechanisms, and so on), 
the results of the study could pose a unique challenge to, for instance, 
numerical simulations of the dynamo mechanisms inside the convective core of early-type main-sequence stars. 

\acknowledgments

We would like to express our gratitude to the NASA and \textit{Kepler} team for the precious data. 
J. Jackiewicz is thanked for his constructive comments. 
\edit1{We also would like to thank the anonymous referee for his or her constructive comments.} 
This work was supported by JSPS Grant-in-Aid for JSPS Research Fellow Grant Number JP20J15226. 

\end{document}